# J-band Red Giant Branch Tip Magnitudes for the Distance Indicator

By Hyun-chul Lee (이현철), Cedar Garcia, Ivana Pena

1201 W. University Dr. Edinburg, TX 78539
The University of Texas Rio Grande Valley
Department of Physics and Astronomy

The tip of the red giant branch is one of the widely used distance measurement methods for the relatively nearby galaxies where the bright individual stars are resolved (e.g., Lee, Freedman, & Madore 1993).  Most of the earlier works have used the photometry in the I-band such as HST ACS/WFC F814W or F850LP for the RGB tip method (e.g., McQuinn et al. 2017; Lee & Jang 2017).  Here we look into the RGB tip magnitudes in J-band such as HST WFC3/IR F110W and JWST NIRCam/WF F115W for a wide range of age and metallicity.

Figure 1 depicts that the J-band (HST WFC3/IR F110W and JWST NIRCam/WF F115W) RGB tip magnitudes in the lower panel do not significantly change compared to that of the I-band (HST ACS/WFC F814W, F850LP) in the upper panel for a wide range of metallicity at 13 Gyr.  We have employed the MIST isochrones (Choi et al. 2016) in this work.  We, however, find that the latest PARSEC isochrones (Marigo et al. 2017) show the similar trend.  Moreover, HST WFC3/IR F110W RGB tip magnitudes stay constant within 0.005 magnitudes for stellar populations with old ages (age > 5 Gyr) at given metallicity (Figures 2 and 3).

Once successfully calibrated, the J-band RGBtip magnitudes would serve a better way to measure the distance to the nearby galaxies compared to the traditional I-band RGBtip magnitudes that demonstrate some significant color-dependency (e.g., Figure 15 of Jang & Lee 2017).  It may further ameliorate the disagreement of Hubble constant measurement between the CMB observations and the galactic distances (Freedman 2017).  Moreover, the theoretical study of the color transformation and the constitutional physics on the tip of the red giant branch luminosity (e.g., Serenelli et al. 2017) will enhance the usefulness of the J-band RGBtip magnitudes method for the robust distance measurements of the galaxies.

## References

Choi, J., Dotter, A., Conroy, C., Cantiello, M., Paxton, B., & Johnson, B.D. 2016, ApJ, 823, 102
Freedman, W.L. 2017, Nature Astronomy, 1, 169
Jang, I.S., & Lee, M.G. 2017, ApJ, 835, 28
Lee, M.G., Freedman, W.L., & Madore, B.F. 1993, ApJ, 417, 553
Lee, M.G., & Jang, I.S. 2017, ApJ, 841, 23
Marigo, P., Girardi, L., Bressan, A., Rosenfield, P., et al. 2017, ApJ, 835, 77
McQuinn, K.B.W., Skillman, E.D., Dolphin, A.E., Berg, D., & Kennicutt, R. 2017, AJ, 154, 51
Serenelli, A., Weiss, A., Cassisi, S., Salaris, M., & Pietrinferni, A. 2017, A&A 606, A33

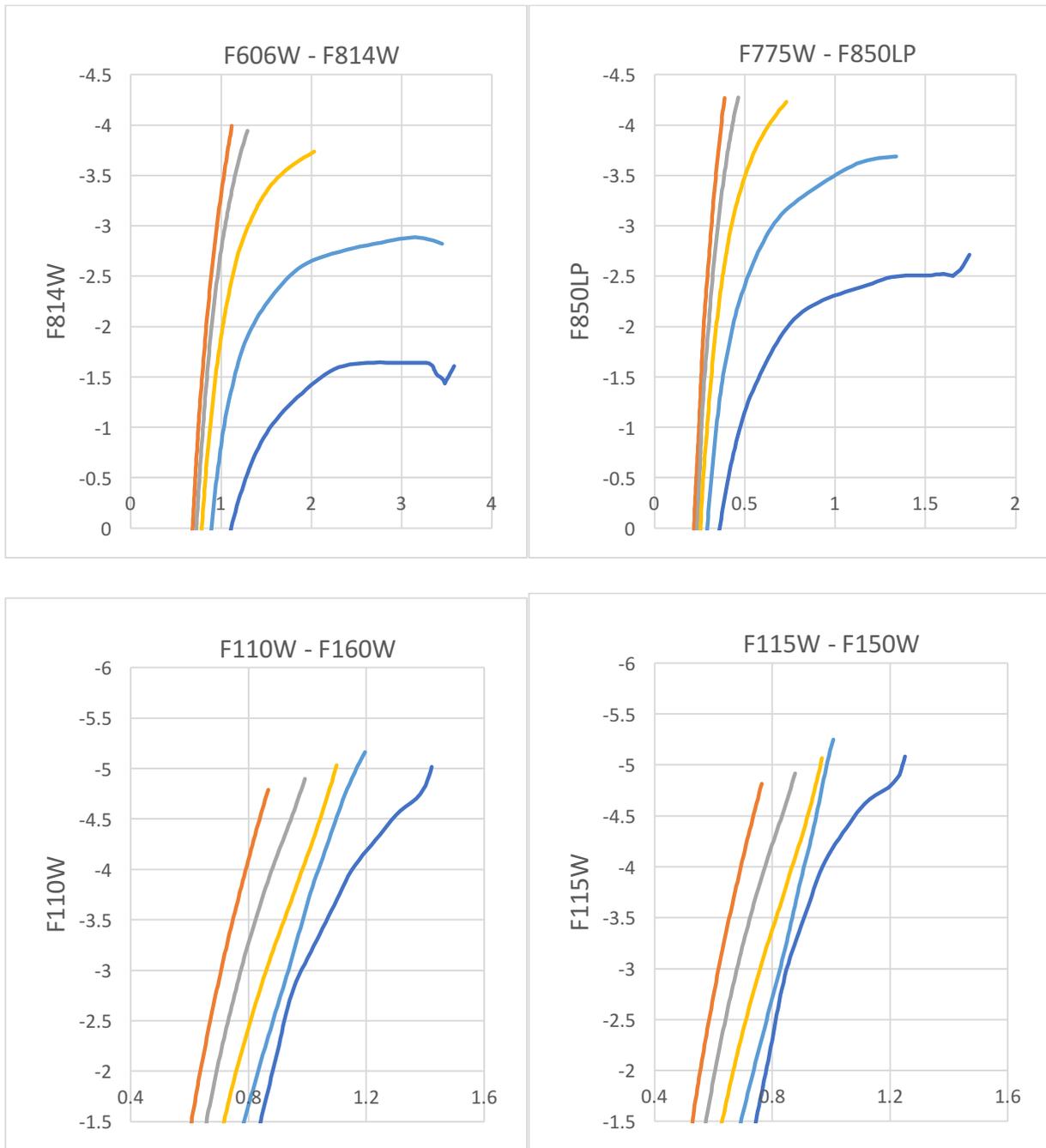

Figure 1. Illustration of the RGB tip magnitudes using the MIST isochrones at 13 Gyr. Upper left: HST ACS/WFC F606W – F814W vs. F814W, upper right: HST ACS/WFC F775W – F850LP vs. F850LP, lower left: HST WFC3/IR F110W – F160W vs. F110W, lower right: JWST NIRCam/WF F115W – F150W vs. F115W. At each color-magnitude diagram, [Fe/H] = -1.5, -1.0, -0.5, 0.0, 0.5 from left to right.

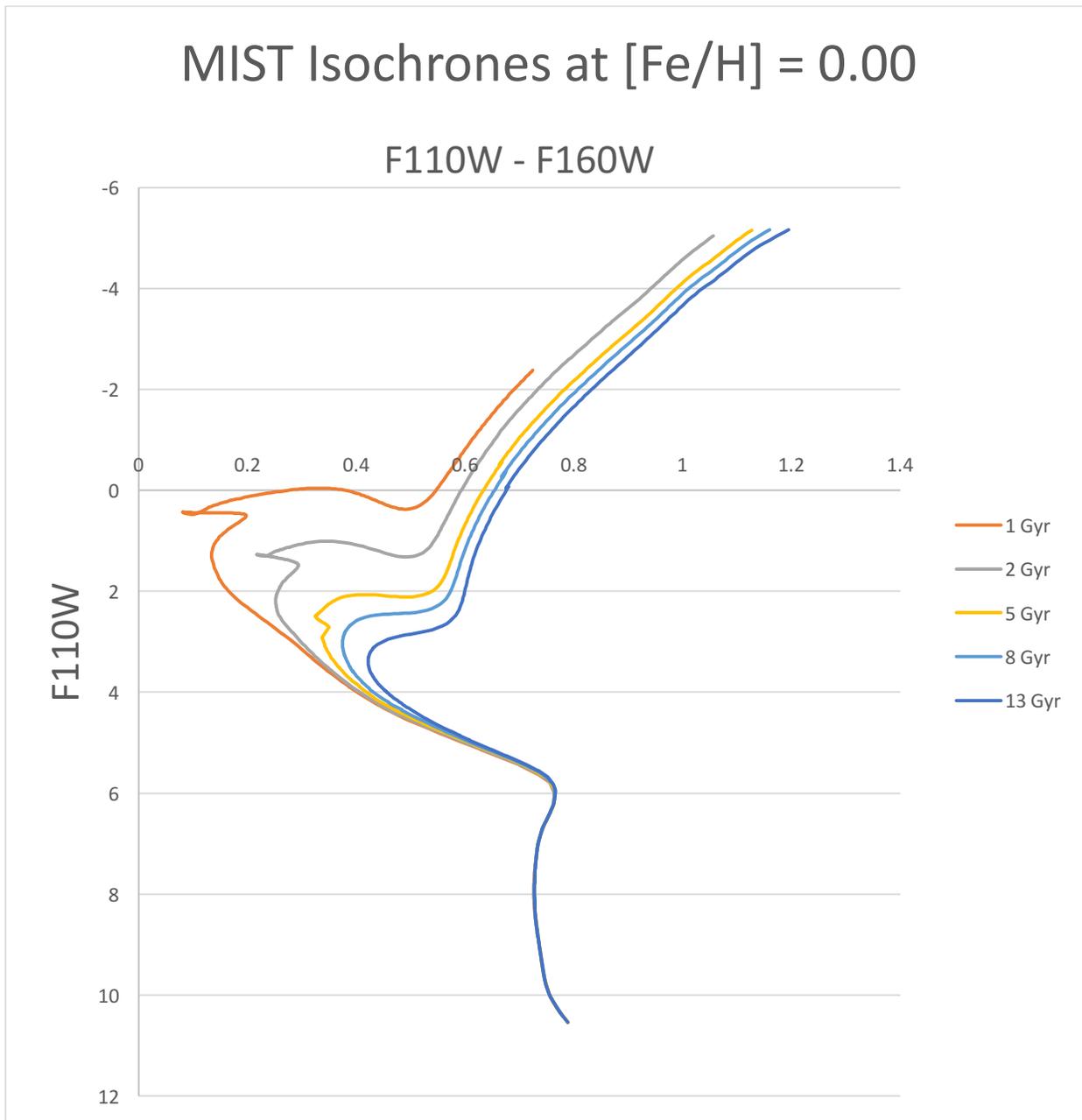

Figure 2. Illustration of the RGB tip magnitudes using the MIST isochrones at [Fe/H] = 0.00 in the HST WFC3/IR F110W – F160W vs. F110W Color – Magnitude Diagram. This Figure is not shown in the RNAAS.

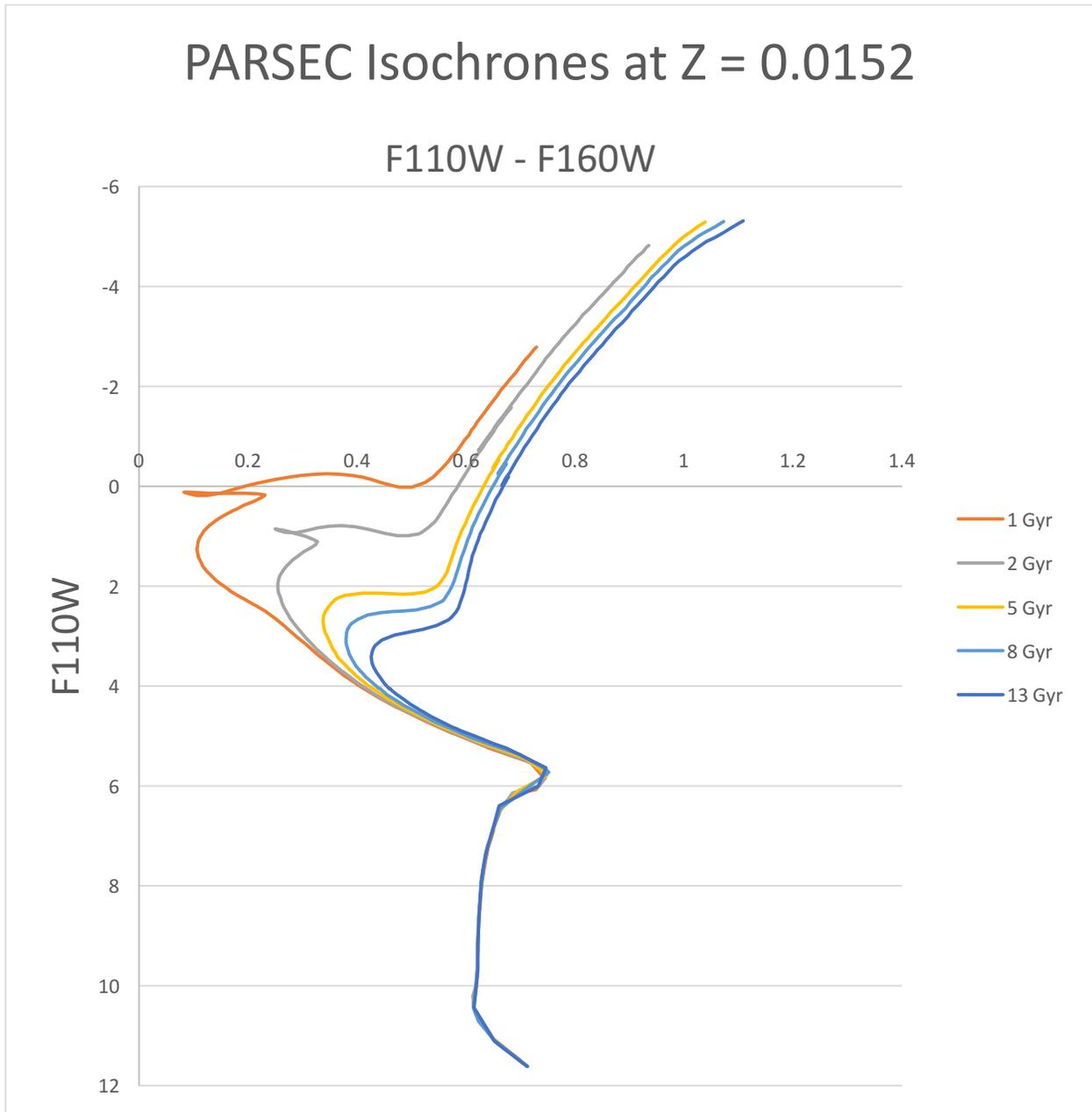

Figure 3. Illustration of the RGB tip magnitudes using the PARSEC isochrones at Z = 0.0152 in the HST WFC3/IR F110W – F160W vs. F110W Color – Magnitude Diagram. This Figure is not shown in the RNAAS.